\documentclass{emulateapj} 

 \usepackage{graphicx}
 \usepackage{epstopdf}
 \usepackage{subfigure}
 
\received{} 
\accepted{} 
\slugcomment{Draft: Submitted to ApJ Supp.} 
\newcommand{\ms}{\mbox{m s$^{-1}~$}} 
\newcommand{\mse}{\mbox{m s$^{-1}$}} 
\newcommand{\kms}{\mbox{km s$^{-1}~$}} 
\newcommand{\kmse}{\mbox{km s$^{-1}$}}

\newcommand{\alln}{2046 }
\newcommand{\stann}{131 }
 
\shortauthors{Chubak{\it et~al.\/}} 
\shorttitle{Velocities of  \alln Nearby FGKM Stars} 

\begin{document} 
 
\title{Precise Radial Velocities of \alln Nearby FGKM Stars  and \stann Standards\altaffilmark{1}}
  
\author{Carly Chubak\altaffilmark{2}, 
Geoffrey W. Marcy\altaffilmark{2}, 
Debra A.\ Fischer\altaffilmark{5},
Andrew W. Howard\altaffilmark{2,3},\\
Howard Isaacson\altaffilmark{2}, 
John Asher Johnson\altaffilmark{4},
Jason T.\ Wright\altaffilmark{6,7}\\}
 
\email{gmarcy@berkeley.edu} 
 
\altaffiltext{1}{ Based on observations obtained at the 
W.M. Keck Observatory, which is operated jointly by the 
University of California and the California Institute of Technology, 
and on observations obtained at the Lick Observatory which is 
operated by the University of California.} 
 
\altaffiltext{2}{ Department of Astronomy, University of California, 
Berkeley, CA USA 94720} 
 \altaffiltext{3}{ Townes Fellow, Space Sciences Laboratory, University of California, Berkeley, CA 94720, USA} 
 \altaffiltext{4}{ Department of Astrophysics, California Institute of Technology, Pasadena, CA 91125, USA} 
\altaffiltext{5}{Department of Astronomy, Yale University, New Haven, CT 06511, USA } 
 \altaffiltext{6}{ Department of Astronomy \& Astrophysics, The
   Pennsylvania State University, University Park, PA 16802, USA}  
 \altaffiltext{7}{Center for Exoplanets and Habitable Worlds, The Pennsylvania State University, University Park, PA 16802, USA} 
  
\begin{abstract} 
We present radial velocities with an accuracy of 0.1 \kms for \alln
stars of spectral type F,G,K, and M, based on $\sim$29000 spectra
taken with the Keck I telescope.  We also present \stann FGKM standard
stars, all of which exhibit constant radial velocity for at least 10
years, with an RMS less than 0.03 \kmse.  All velocities are measured
relative to the solar system barycenter.  Spectra of the Sun and of
asteroids pin the zero-point of our velocities, yielding a velocity
accuracy of 0.01 \kmse for G2V stars.  This velocity zero-point agrees
within 0.01 \kms with the zero-points carefully determined by
\cite{Nidev} and \cite{Latham02}.  For reference we compute the
differences in velocity zero-points between our velocities and
standard stars of the IAU, the Harvard-Smithsonian Center for
Astrophysics, and l'Observatoire de Geneve, finding agreement with all
of them at the level of 0.1 \kmse.  But our radial velocities (and
those of all other groups) contain no corrections for convective
blueshift or gravitational redshifts (except for G2V stars), leaving
them vulnerable to systematic errors of $\sim$0.2 \kms for K dwarfs
and $\sim$0.3 \kms for M dwarfs due to subphotospheric convection, for
which we offer velocity corrections.  The velocities here thus
represent accurately the radial component of each star's velocity
vector.  The radial velocity standards presented here are designed to be
useful as fundamental standards in astronomy. They may be useful
for Gaia \citep{Crifo10, Gilmore12} and for dynamical studies of such systems
as long-period binary stars, star clusters, Galactic structure, and
nearby galaxies, as will be carried out by {\em SDSS, RAVE, APOGEE, SkyMapper, HERMES, and LSST}.
\end{abstract}


\section{Introduction} 
\label{intro} 

Doppler shifts of stellar spectra provide information about the
line-of-sight component of the velocity vector of the target stars in
the frame of the telescope.  When transformed to the frame of the
center of mass of the Solar System, those "barycentric" radial
velocities represent the star's velocity component measured relative
to a well defined, and commonly adopted, inertial frame within our
Milky Way Galaxy, suitable for studying the motions of stars in a wide
variety of astronomical settings.

Barycentric radial velocities enable study of the kinematics,
structure, and mass distribution of the Milky Way Galaxy, including
the disk components, bulge, nucleus, and halo.  Doppler measurements
also provide a primary tool for detecting and characterizing binary
stars in many environments such as in the general field, open
clusters, star forming regions, planet-hosting stars, globular
clusters, and the Galactic center.  Radial velocities also serve to
measure the dynamics and mass content, both luminous and dark, of star
clusters and galaxies.  Moreover, radial velocities are vital for
measuring the infalling extragalactic matter into our Galaxy, as well
as the dynamically important motions of stars within other galaxies in
the local group.

When combined with the proper motion and positions measured using,
e.g., the Hipparcos telescope or the upcoming Gaia space telescope,
one can measure the three dimensional velocity vectors of stars and
stellar systems.  Such three dimensional velocity measurements offer
information about the origin, history, future, and mass distribution
of the components of the Galaxy.  Precise radial velocities measured
over time can reveal the acceleration of stars, caused surely by
gravitational forces exerted by unseen nearby objects including
orbiting planets, brown dwarfs, and stars, as well as by nearby
compact objects such as white dwarfs, neutron stars, and black holes.

Many new observational facilities are now, or soon will be, providing
kinematic and positional information about stars in the Galaxy.  The
{\em RAdial Velocity Experiment} (RAVE) is studying the properties and
origin of the structure in the Galactic disc \citep{Wilson2011} by
measuring velocities with an accuracy of $\sim$2 \kms for up to
500,000 stars.  The Sloan Digital Sky Survey and the {\em Sloan
  Extension for Galactic Understanding and Exploration} (SEGUE), with its
imaging and radial velocity capability (with 10 \kms accuracy) are
performing extraordinary measurements on the kinematics of the Galaxy
and its halo \citep{Schoenrich2010}.  Some groups are combining
spectroscopy with these kinematic measurements to gain unprecedented
information about the coupled chemical and kinematic properties of the
solar neighborhood in the Galactic context \citep{Casagrande2011}.
The Gaia-ESO Survey with VLT/Flames will be particularly valuable with its
spectroscopy of 100,000 stars in all stellar populations and will nicely
complementing the measurements of Gaia \citep{Gilmore12}.

Radial velocity standard stars of a wide range of stellar types
provide useful Doppler calibrators for the many different instruments
doing this kinematic work.  The radial velocity standard stars provide
touchstones of comparions for both zero-points and the velocity scales
of the different instruments and observatories.  

Despite these current and future uses of radial velocities, one might
wonder whether highly precise, absolute radial velocities have value in the modern era.  After all, it is
relative velocities not absolute barycentric velocities that are used
to discover orbiting exoplanets, measure orbits of binary stars,
measure velocity dispersions in virialized systems of stars, and to
measure masses of compact objects including supermassive black holes.
Moreover accurate radial velocities could be obtained by using
carefully chosen reference template spectra, either observed (of
asteroids, for example) or synthetic.  These arguments for the
obsolescence of absolute velocities may have some validity for those
specialists performing radial velocity measurements.  But for the
majority of astronomers there remains a widespread need to measure
radial velocities as part of a larger project for which exhaustive
calibration is not practical.  Radial velocity standards offer a sound
method at the telescope to place one's Doppler measurements on a well
established scale and to learn their accuracy by observing multiple
standard stars with one's particular (and sometimes peculiar)
instrument.  Further, new spectrometers on the ground or in space
often come online with uncertain wavelength scales, structural and
thermal instabilities, or errors that depend on stellar temperature, all of
which can be mitigated by a real-time determination of the velocity
zero-point and scale.  Radial velocity standard stars offer that
metric to establish and demonstrate accuracy and internal precision.

Some information about radial velocity standards is maintained by the International
Astronomical Union (IAU), Commission 30 found at \verb|http://sb9.astro.ulb.ac.be/iauc30/.|
The IAU  has constructed a precise definition of ``radial velocity''
described by \cite{Lindegren03}.

The old standard source of radial velocities was the General Catalog
of Stellar Radial Velocities (BCRV) prepared at the Mount Wilson
Observatory in Pasadena, California \citep{Wilson53}.  The modern era
of radial velocity standards occurred with the work by the Geneva
group led by Michel Mayor and Stephane Udry, the University of
Victoria group led by Collin Scarfe and Robert McClure, and the group
at the Harvard-Smithsonian Center for Astrophysics led by David Latham
and Robert Stefanik \citep{Mayor85, Scarfe90, Latham91, Latham02}.  An excellent
summary of the history of radial velocity standards through 1999 is
provided by \cite{Stef99}.

Three excellent radial velocity programs provided standards with high
accuracy and integrity \citep{Stef99, Udry99a, Udry99b}, all of them
constituting a modern velocity zero-point with accuracy better than
0.3 \kmse.  Additional excellent stellar radial velocity measurements
were made by \cite{Nordstrom04, Famaey05} at accuracies of $\sim$0.3
\kmse, and also by the Fick Observatory at Iowa State University, and
at the Mt. John University Observatory in Christchurch, New Zealand
\citep{Beavers86, Hearnshaw99}.  The Pulkova Radial Velocity Catalog
compiles the mean velocities for over 35000 Hipparcos stars
\citep{Gontcharov06}.  The velocities come from over 200 publications,
yielding a median accuracy of 0.7 \kms.

The largest modern catalog of radial velocity standard stars was
established by \cite{Nidev} who measured the radial velocities of 889
FGKM-type stars with an accuracy of 0.1 \kms in the solar system
barycentric frame.  \cite{Nidev} made multiple radial velocity
measurements with the Keck 1 telescope and High Resolution Echelle
Spectrometer (Vogt et al. 1994) over a typical time span of 3-7
years, thereby revealing any velocity variability.  Use of an iodine
cell to impose a wavelength scale yielded relative radial velocities
with a precision of 0.003 \kms (RMS, with no zero-point) able to
detect tiny velocity variability at the level of 0.01 \kms on time
scales of years.  Thus, the radial velocity standard stars in
\cite{Nidev} met the highest standard for constancy in velocity.  We
adhere to that standard here.

\cite{Nidev} contained more stellar radial velocities at the highest
viable accuracy of 0.1 \kms than any prevous radial velocity survey,
and they provided statistically robust comparisons of the zero-points
of other radial velocity surveys.  The radial velocity measurements of
\cite{Nidev} thus uphold high integrity for a wide range of spectral
types and provide standard stars at all RA and northward of
declination -30 deg.

Here we extend the work of \cite{Nidev} with 9 additional years of
radial velocity measurements from the same Keck 1 telescope and
spectrometer.  These measurements supersede those in the Nidever et
al. paper by providing more velocity measurements over a longer time
baseline, and we include more stars.  We include only those \alln
stars for which we obtained Keck-HIRES spectra using the modern CCD
detector in HIRES that was installed in June 2004, as we use only the
near-IR portion of the spectrum made available at that time.  Thus
some stars listed in Nidever et al. are not included here.  We
establish \stann standard stars, with radial velocities measured
relative to the barycenter of the solar system for spectral types FGKM that are stable at the level
of 0.01 \kms during a decade, and we provide barycentric velocities
for \alln FGKM stars.  The final velocities have an precision of
typically better than 0.1 \kmse, and they reside on the Nidever et
al. velocity scale.  We compare these velocities to extant velocities
by other groups.

\section{Spectroscopic Observations and Velocity Measurements} 

We obtained spectra using the HIRES echelle spectrometer on the 10-m
Keck 1 telescope between 2004 August and 2011 January, as part of the California Planet Survey (CPS) to detect
exoplanets by the Doppler technique, using iodine to calibrate both
wavelength and the instrumental profile spectrometer at each
wavelength \citep{Marcy08, Johnson11}.  Before each observing night we
positioned the CCD so that the (observatory-frame) wavelengths land on
the same pixels within 1/2 pixel as on all previous observing nights.
This produced a nearly identical wavelength scale on all nights.  At
the beginning and end of each observing night, we took spectra of a
thorium-argon lamp, providing the linear and non-linear portion of the
wavelength scale information, but not the zero-point of the wavelength
scale which was instead established by absorption lines formed in the
Earth's atmosphere (see below).  We found that the wavelength
dispersion of HIRES varies by about 1 part in 2000 over the course of
months and years, presumably due to slow mechanical and thermal
changes in the spectrometer optics and to changes in air pressure.  All 29000 spectra of the \alln
stars and all calibration spectra are available on the
Keck Observatory Archive (after the nominal proprietary period of 18 months), made possible by a NASA-funded
collaboration between the NASA Exoplanet Science Institute (NExSci) and the
W. M. Keck Observatory\footnote{http://ssd.jpl.nasa.gov/}.

We employed an exposure meter to set exposure times
for each observation, promoting uniform and high signal-to-noise
spectra during the seven years of observations \citep{Kibrick06}.  The spectra had a
typical S/N~$\sim$~150 per pixel at 720 nm, which is near the center
of the near-IR wavelength domain, 654 - 800 nm, used in this paper.
The spectral resolution was, R~=~55,000 and the pixel spacing
corresponds to a Doppler shift of 1.3 km s$^{-1}$ per pixel at the
blaze wavelength of all spectral orders.  The dispersion was found to
vary by $\sim$10\% along the free spectral range of each order.  The
spectrometer instrumental profile had a typical FWHM of $\sim$4.2 pixels, varying by
10-20\% as a function of wavelength.  In this current work, we used
both types of spectra obtained as part of the CPS planet-search
program, namely those with the iodine cell in front of the
spectrometer slit for which the starlight passed through the molecular
iodine gas ~\citep{Marcy92} and those with the iodine cell not in the beam that
contain no iodine lines.  The presence of iodine cell has little
effect on the Doppler measurements because the iodine lines are less
than 1\% deep in the near-IR wavelength regions used here.

We carefully determined a wavelength scale for each spectrum.  We
first fit a fifth-order polynomial to the positions of the thorium
lines, with their associated wavelengths, to determine a
first-approximation to the wavelength scale (called "thid" files).  We
used a spline to map the spectra onto a logarithmic wavelength scale,
based on the original wavelength scale from the thorium-argon spectra.
The new array pixels were designed to be separated by equal intervals
in delta $\ln \lambda$.  This logarithmic wavelength scale offers the
advantage that a certain difference in radial velocity, $\Delta v$,
causes the spectrum to be displaced by the same distance in units of
pixels at all wavelengths, given by $\Delta v = c\Delta \ln{\lambda}$.
The typical pixel size for a Keck HIRES spectrum corresponds to a
Doppler shift of 1.3 \kmse, and this thorium-based wavelength scale
determined the {\em relative} wavelength scale to 6 significant digits
($\sim$ 0.1 pixel), based on the scatter in the fits.  The wavelength
zero-point remained to be set accurately (using telluric lines, as
described below).  Instead of cross-correlation, we used the
chi-square statistic to determine the relative shifts between a
template spectrum and observed spectrum.  To measure fractional pixel
shifts, we oversampled, at 0.01 \kms per pixel, a subarray around the minimum of each
chi-squared function and interpolated with a spline function.

\begin{figure} 
\includegraphics[scale = 0.6]{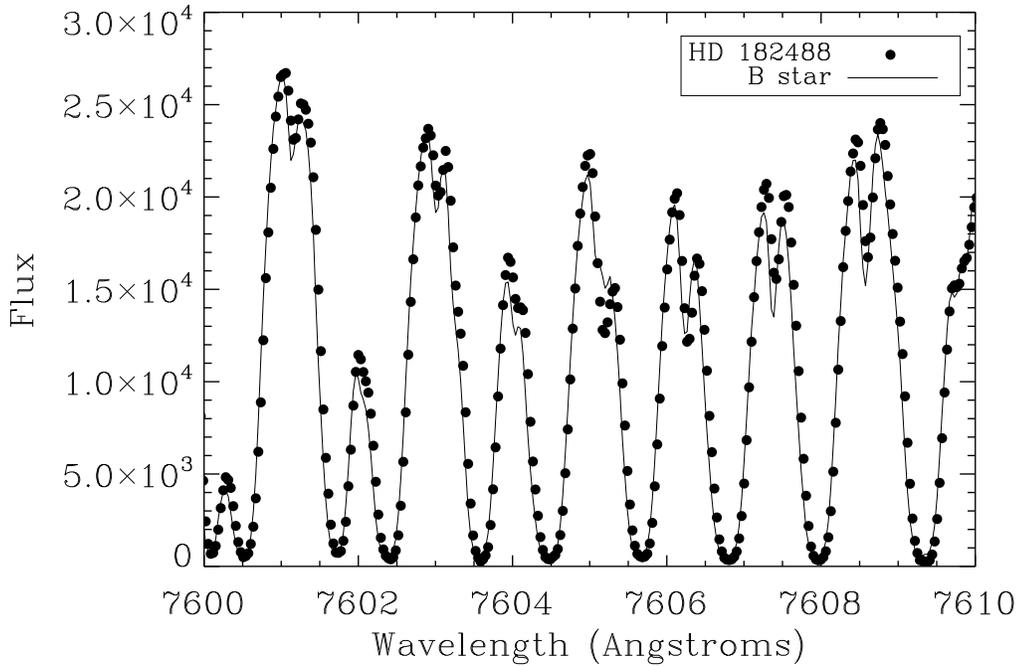} 
\caption{Telluric A band seen in both the reference B star, HD 79439 (solid line) and the
program star, HD 182488 (dots).  Careful inspection shows that the
telluric lines in the spectrum of the program star are slightly
displaced redward of those in the B star by 0.437 pixels, due
primarily to a difference in the zero-points of the two wavelength scales caused by instrumental effects.  We remove these differences in wavelength zero-point to within a hundredth of a pixel by $\chi^2$ fitting of one spectrum to the other, with the displacement the free parameter.   This yields final radial velocities precise to better than 0.1 \kmse.} 
\label{fig:compare_telluric} 
\end{figure}

To determine a secure wavelength zero-point for each spectrum, we
followed the suggestion of \cite{Griffin73} by using telluric lines to
determine the wavelength scale. We used the telluric A and B
absorption bands, at wavelengths of 759.4-762.1 nm \ and 686.7-688.4
nm \ respectively, due to absorption by molecular oxygen in the
Earth's atmosphere.  We again used a $\chi^2$ minimization method
to find the displacement of the telluric lines in the program star
relative to those in the reference B-type star spectrum of HD 79439.
Figure \ref{fig:compare_telluric} shows the telluric lines in both
that reference spectrum and in the representative spectrum of a
program star, HD 182488.  The program star spectrum has telluric lines
clearly displaced by a fraction of a pixel (redward), in this case by
+0.437 pixels, an amount representative of the typical shifts in
wavelength zero-point from observation to observation over the time
scale of days, months, and years covered by the spectra presented
here.  This measureable displacement of telluric lines provides the
key correction to the wavelength zero-point that accounts for small
changes (typically less than 1 \kmse) in the CCD position, the
spectrometer optics, and (importantly) for the non-uniform
illumination of the starlight on the entrance slit of the
spectrometer.

This approach corrects for the dominant systematic errors that often
compromise normal radial velocity measurements that do not use an
absorbing gas to establish the wavelength scale of a slit-fed (rather
than fiber-fed) spectrometer.  We subtracted this displacement from
the apparent shift of the stellar lines in order to find the net
(true) Doppler shift. In this way we found the radial velocity
determined by the Doppler shift of stellar absorption lines, including
a correction for the shift in the wavelength zero-point. Using
telluric lines to set the wavelength zero-point leaves systematic
errors of $\sim$0.01 \kms caused by typical winds in the Earth's
atmosphere.  Also, the telluric lines we used (the A and B bands) are
not distributed in wavelength coincident with the stellar lines we
employed.  An additional Doppler error may accrue due to unaccounted
for nonlinear errors in the wavelength scale.  We find such errors to
be several tens of meters per second.

We chose four wavelength segments in the near-IR that are rich in
stellar absorption lines in FGKM stars, but nearly free of telluric
lines.  Using a spectrum of the A5 star, HR 3662, we determined which
regions of the spectra in the near-IR of HIRES spectra are largely
unpolluted by telluric lines.  The resulting four wavelength regions
were 679.5-686.7 nm, 706.7-714.6 nm, 739.8-748.9 nm, and 751.8-759.3
nm.  In these wavelength regions, we carried out Doppler measurements
by using standard chi-square minimization as a function of Doppler
shift between the spectra of the program star and template reference
star.  We averaged the velocities from the four wavelength segments of
the spectrum, and we recorded that average value as the star's radial
velocity.  We employed two template spectra.  For the FGK stars, we
used a spectrum of sun-light reflected off Vesta as a solar proxy.  As Vesta is unresolved,
its use ensured that the spectrometer optics were illuminated in
nearly the same way as by the program stars.  For M dwarfs we used a
spectrum of HIP 80824 (spectral type M3.5).  We applied a barycentric
correction to the velocity for each spectrum in the frame of the solar
system barycenter.
 
 The resulting "raw" radial velocities were systematically different
 from those of \cite{Nidev} by an arbitrary constant amount due to the
 radial velocity and barycentric correction of the template spectrum.
 We calculated this constant by taking a sample of 110 standard FGKM
 stars in \cite{Nidev} and comparing those velocities to our measured
 raw velocities for those stars.  We determined the average difference
 between our raw velocities and those of \cite{Nidev}, constituting
 the constant to be applied to all of our velocity measurements.  This
 automatically forces our radial velocities to have the same
 zero-point as those of \cite{Nidev}.

Thus our velocity measurements reside on the scale of \cite{Nidev} by
construction.  These radial velocities of stars are measured relative
to a hypothetical inertial frame located at the barycenter of the
Solar System.  This transformation is accomplished by using the JPL
ephemeris of the Solar System to determine the velocity vector of the
Keck 1 telescope at the instant of the photon-weighted midpoint of the
exposure of the spectrum (accurate to within a few seconds).  Our
transformation to the barycentric frame is performed using the JPL
ephemeris\footnote{http://ssd.jpl.nasa.gov/}, accessed and interpreted
with utilities from the IDL Astronomy User's
Library\footnote{http://idlastro.gsfc.nasa.gov/} and custom driver
codes written by the California Planet Survey.  We carried out
extensive tests of our barycentric transformation code, finding
discrepancies of 0.1 \ms  in comparison with
TEMPO 1.1 which is similar to that of TEMPO 2 \citep{Edwards2006}.
Errors of that magnitude, 10$^{-4}$ \kmse, are negligible compared to
other sources of error in this present work.

As usual for such transformations to the Solar System barycenter, we
do not include the effects of the solar gravitational potential at
that location (near the surface of the Sun) that would cause a
(meaningless) gravitational blueshift.  Similarly, we do not account
for the gravitational blueshift caused by starlight falling into the
potential well of the Sun at the location of the Earth, a $\sim$3 \ms
effect.  We also do not take into account the gravitational redshift
as light departs the photosphere of the star, an effect of hundreds of
\ms that depends on stellar mass and radius.  

We further ignore the convective blueshift of the starlight caused by
the Doppler asymmetry between the upwelling hot gas and the
downflowing cool gas.  Convective blueshift depends on spectral type
\citep{Dravins}, and we do not include any theoretical estimates of
this photospheric hydrodynamic effect here.  Both gravitational
redshift and convective blueshift amount to a few tenths of a
kilometer per second, and while they are opposite in sign they may not
cancel each other.  However see Section 4 for a quantitative
discussion of these two effects, that appear to largely cancel each
other.  We note that several efforts have been successful at measuring
the convective blueshift in a few stars, especially for the Sun and
the alpha Centauri system \citep{Ramirez10, Dravins08, Nordlund09,
  Pourbaix02}.

\subsection{Radial Velocity Standard Stars}

\begin{figure} 
\includegraphics[scale = 0.6]{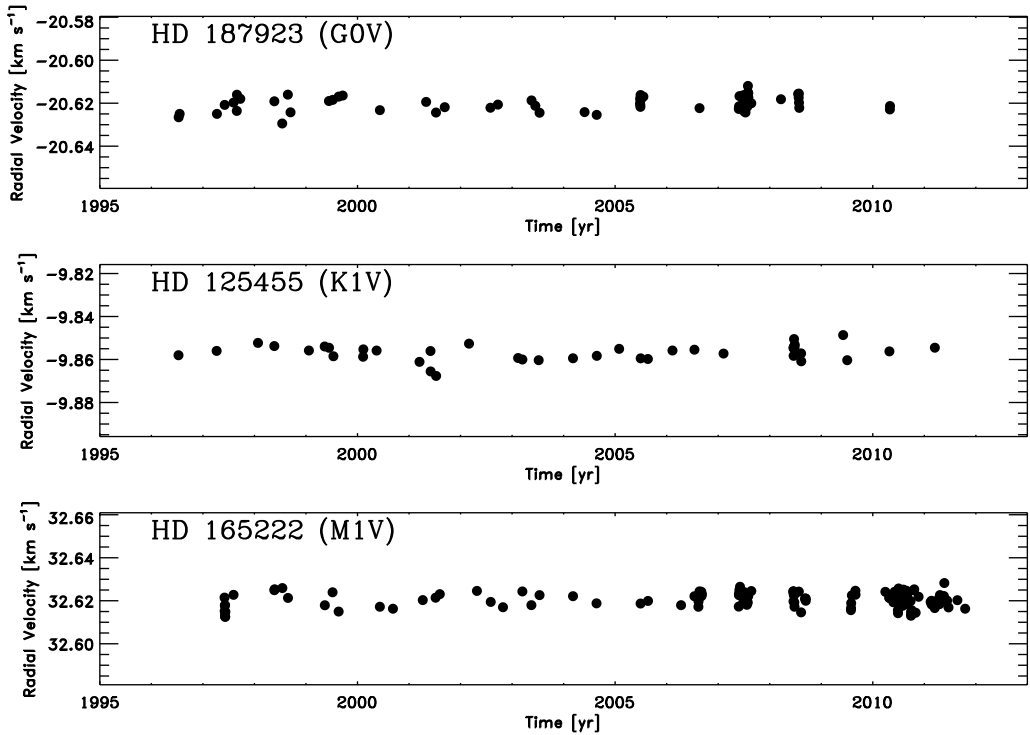} 
\caption{Precise measurements of barycentric radial velocity vs. time for three representative standard stars listed in Tables 1 and 2.  These relative radial velocities are measured using iodine gas in absorption at the telescope, providing the wavelength scale superimposed directly on the stellar lines \citep{Marcy92}, and yield a typical precision of 0.002 \kmse (2 \mse, RMS).  The zero-point of these velocities is determined using the present technique that employs telluric lines, and the relative velocities come from the iodine-based measurements.  The three standard stars exhibit velocity variation of less than 0.020 \kms (RMS) during a time span of over 10 years, meeting the criteria for status as a standard star in this paper.} 
\label{fig:prv_standard} 
\end{figure}

We identified standard stars from among the \alln total sample of
stars based on several criteria.  We examined the iodine-based
relative velocities, having a precision of $\sim$3 \mse, for each of
the \alln stars.  We established a severe criterion of stability
during 10 years in order for a star to be qualified as a "radial
velocity standard star".  All standard stars must exhibit an RMS of
their iodine-based relative velocities under 0.03 \kms \citep{Marcy92}
and a duration of such velocity measurements of at least 10 years.
Figure \ref{fig:prv_standard} displays the iodine-based relative
velocities for three representative standard stars.  The absolute
radial velocity relative to the Solar System sets the zero-point and
the relative velocities come from the iodine-based Doppler
measurements.  The velocities of the three representative cases
exhibit an RMS of under 0.010 \kms (10 \mse) and span over 10 years,
typical of the standard stars, promoting their integrity as standard
stars at the more relaxed level of 0.1 \kmse.

Thus, our radial velocity standard stars must demonstrate constant
velocity during a decade within a tolerance of 30 \ms RMS during that
time.  We required that at least 3 spectra be obtained over a 10 year
time period to demonstrate the decade-long stability.  We identified
\stann standard stars based on these criteria.  Among them, only 12
exhibit an RMS scatter in their iodine-based RVs of more than 0.01
\kmse, and none over 0.03 \kmse, during 10 years of observations.
Thus, the \stann standard stars all exhibit radial velocity stability
during a decade at the level of 0.03 \kmse.

The barycentric radial velocities for the \stann standard stars are
reported in Tables 1 and 2.  In Table 1, primary and alternate star
names are given in the first three columns, and the spectral type is
given in column 4.  Columns 5, 6, and 7 give the Julian dates of the
first and last observations, and the duration of observations in
years.  Column 8 lists the radial velocity of the star relative to the
solar system barycenter given by \cite{Nidev}.  Column 9 lists the
unweighted average of all radial velocity measurements from this
current work.  In the next two columns we list the standard deviation
of the multiple velocities we measured here for the star and the
number of spectra used.  We report the final velocity for each
standard star as the average of the \cite{Nidev} and present
velocities.  We consider this final radial velocity to be robust as
both sets of velocities have high integrity and we do not rank one set
significantly higher in integrity than the other.  Importantly, the
Nidever et al. velocities and these new velocities were determined
using completely different Doppler algorithms and different wavelength
regions.  Thus the Nidever and present radial velocities offer
considerable resistance to unexpected errors associated with any
particular method or wavelength.  The Nidever et al. velocities had a
wavelength scale rooted in the iodine lines and measured using the
500-600 nm wavelength region, quite different from the wavelength
scale here rooted in telluric lines at 670 and 760 nm and measured
using the near IR spectrum.

The radial velocity uncertainty recorded in Tables 1 and 2 is the
largest of three values: the difference between our present velocity
and Nidever's, the uncertainty in the mean, or 0.03 \kms, which we
deemed our base accuracy, to prevent artificially low uncertainties.
Table 2 lists the same standard stars, but in a format more suitable
for observing.  We give the primary name, position in RA and DEC,
magnitude, spectral type, final absolute radial velocity, and
uncertainty.

Establishing and maintaining a single, well-defined velocity scale,
including zero-point accuracy and precision, is important to make the
radial velocities more useful.  The velocity scale must be compared to
other well--known scales.  In particular, our velocities are compared
here to those from Geneva, Harvard-Smithsonian, and the California
Planet Survey\footnote{http://exoplanets.org}.

We compared our "present" velocities to those of the standard stars of
\cite{Udry99a}.  The average
of the differences (i.e. zero--point difference) is: 

$$<V_{\rm present} - V_{\rm Udry}> = +0.063 \ \kmse.$$  

\begin{figure} 
\includegraphics[scale = 0.6]{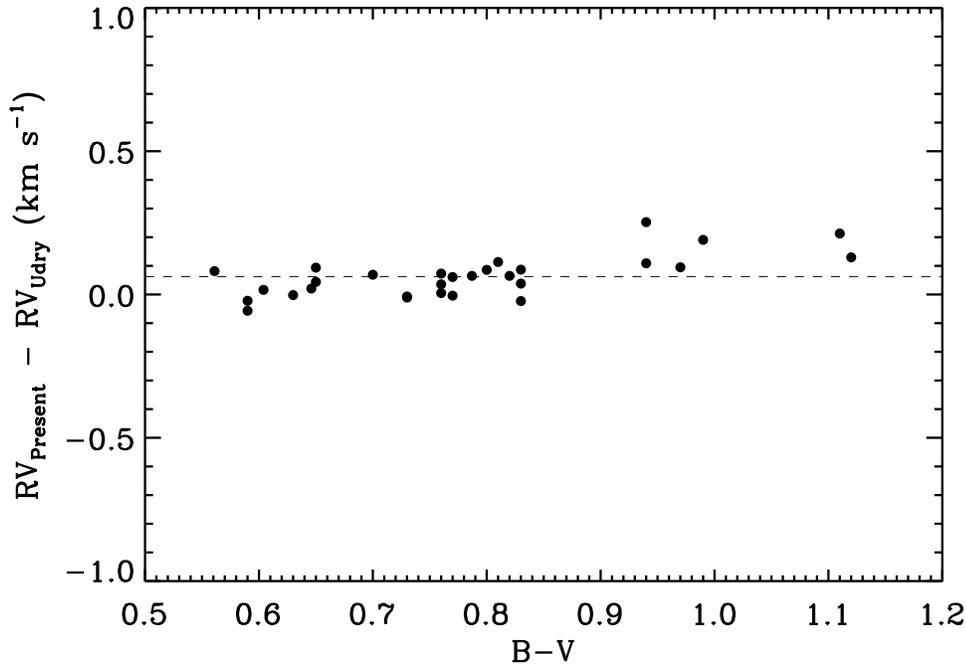} 
\caption{Difference in radial velocities of standard stars between those measured here and those from \cite{Udry99a}, for all stars in common, as a function of B-V color.  The RMS scatter is 0.072 \kmse, and the means of the two sets of radial velocities differ by, (present-Udry) = 0.063 \kmse.   However, there is clear evidence of a systematic trend in the difference in the velocity zero-points with stellar color such that the present-Udry velocities increase by 0.15 \kms between G0 and M0 stars. The origin of this difference is difficult to trace.} 
\label{fig:compare_udry} 
\end{figure}

Thus there is a statistically significant difference in the
zero-points.  That this difference is less than 0.1 \kmse offers some
scale to the integrity of the discrepancies in the two systems of
radial velocities.  The RMS of the differences is 0.072 \kms (RMS) for
the 30 standard stars in common, as shown in Figure
\ref{fig:compare_udry}.  Thus the two sets of velocities agree within
0.1 \kmse in zero point and scatter.  However, the differences in the
velocities appear to be correlated with stellar B-V color, suggesting
a systematic error.  Inspection of Figure \ref{fig:compare_udry} shows
that it is the M dwarfs, with B-V $>$ 0.9 where the systematic
difference resides of (present - Udry) = +0.10 \kmse.  Thus, while the
FGK stars of Udry et al. and the present set have different velocity
zero-points by 0.063 \kmse, the M dwarfs differ by 0.10 \kms.

\begin{figure} 
\includegraphics[scale = 0.6]{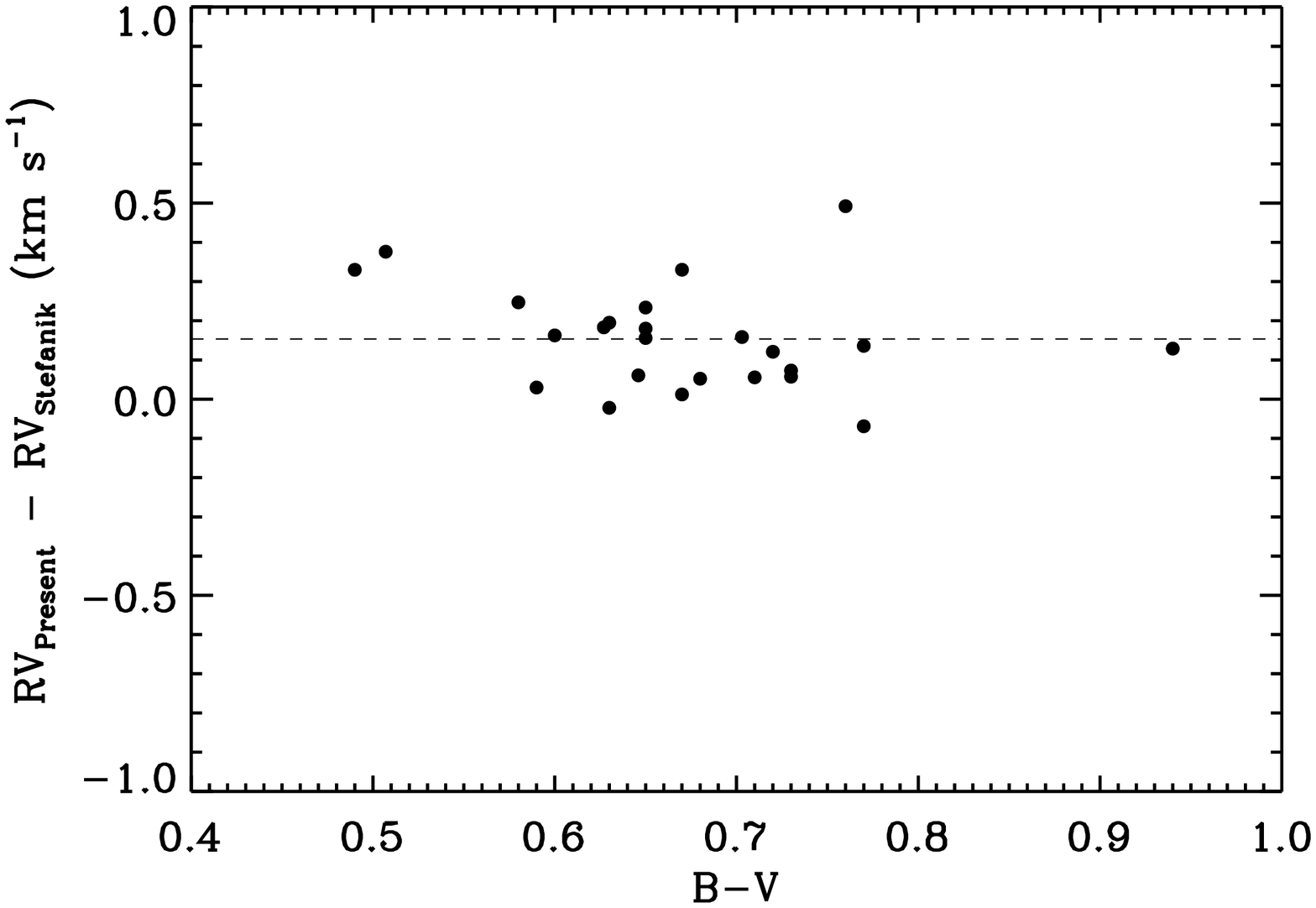} 
\caption{Difference in radial velocities of standard stars between
  those measured here and those from \cite{Stef99}, for 25 standards
  in common. The RMS scatter is
  0.13 \kms and the means of the two sets of radial velocities differ
  by (present-Stefanik) = +0.15 $\pm$ 0.026 \kmse.  Apparently these two sets of
  standard stars have velocities consistent at the level of 0.15
  \kmse.  Doppler measurements
  of asteroids raises the CfA zero-point by 0.139 \kms, bringing the two
zero-points to agreement within 0.01 \kmse.}
\label{fig:compare_stef} 
\end{figure}

We also compared our velocities to those of \cite{Stef99}. 
Considering the 25 standard stars in common, as seen in Figure
  \ref{fig:compare_stef}, the average of the differences is: 

$$(V_{\rm present} - V_{\rm Stefanik}) = +0.15\ \kmse.$$ 

The differences in the 25 velocities have a scatter of 0.13 \kms (RMS).
Thus the present velocities differ in zero-point from those of
\citet{Stef99} by a statistically significant amount (see Section 4 for the explanation).

A useful compilation of velocities was provided by \citet{Crifo10}
based on various past surveys.  They show that the velocities from
Nidever et al. (2002) are valuable because of the accuracy ($<$ 0.1
\kms), the large number of observations, and the long duration of the
velocity time series. They compare the Nidever et al. velocities to
those from past CORAVEL measurements that have typical accuracy of 0.3
\kmse, finding good agreement within errors.  They offer a preliminary
list of standard stars drawn heavily from \citet{Nidev}.  Thus the
zero-point and scale of the velocities in Crifo et al. naturally agree
with those here.

\begin{figure} 
\includegraphics[scale = 0.6]{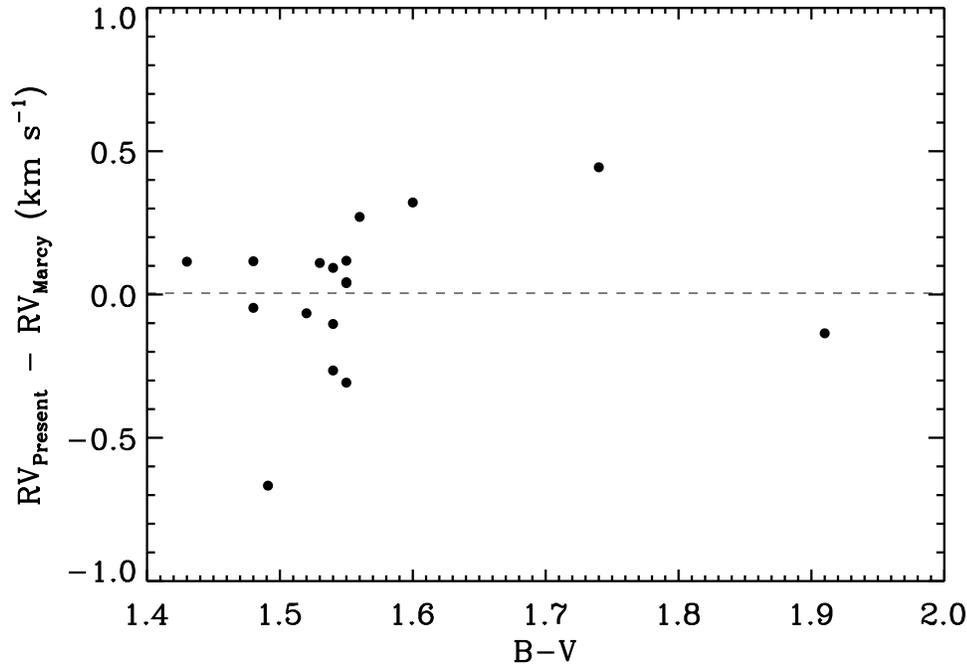} 
\caption{Difference in radial velocities of M dwarfs deemed standard
  stars between the velocities measured here and those from \cite{Marcy}. The differences have an RMS scatter of 0.26 \kms and a zero-point
  difference of +0.007 \kms which is insignificant.  The level of
  agreements between these sets of velocities implies errors of no
  more than 0.26 \kmse, most of which likely comes from velocity errors in \cite{Marcy}}. 
\label{fig:compare_marcy} 
\end{figure}

We also compare our measurements of M dwarfs to those of \cite{Marcy}.
We find that the velocity differences scatter by 0.26 \kms (RMS) and our
zero--points are different by

$$<V_{\rm present} - V_{\rm Marcy}> = 0.007\ \kms$$ 
for the 17 stars in common (see Figure \ref{fig:compare_marcy}).  As the
velocities from \cite{Marcy} are expected to carry precision of only
$\sim$ 0.2 \kmse, this scatter of 0.26 \kms RMS is consistent with
most of the error residing in \cite{Marcy} and only $\sim$0.1 \kms
residing in the errors in the present velocities of M dwarfs.

The differences in the velocities between those of \cite{Nidev}
and those of both \cite{Udry99a}
and \cite{Stef99} exhibted a scatter of less than 0.1 \kms (RMS), and our present
velocities differ from those previous standard measurements within a
margin of 0.1 \kms (RMS).  {\em Thus, the velocities reported here
  agree with the best established standard stars to within 0.1 \kms in
  precision, with modest zero-point differences of comparable
  magnitude.}

We show in section 2.3, that the radial velocities measured
here for 428 stars in common with Nidever et al. (2002) agree within
$\sim$0.13 \kms (RMS) and that there is little dependence on the color
of the stars.  This comparison of 428 stars offers further weight to
the suggestion that the standard stars in Tables 1 and 2 have
integrity at the level of 0.1 \kmse.  We also show in Section 4 that
the zero-point of the velocity scale has integrity at the level of 0.1 \kmse.

\subsection{Uncertainty in the Velocities of Standard Stars}

We compute the uncertainty of the radial velocity for each of the
\stann standard stars by considering two separate estimates of the
uncertainty.  The first estimate is the uncertainty of the mean
velocity measurement, defined as $\sigma/N_{obs}^{1/2}$, where
$\sigma$ is the standard deviation of the ensemble of velocities for a
particular star.  This estimate offers a measure of the internal
uncertainty revealed by the scatter in the individual velocity
measurements.  As a second estimate of uncertainty we compute the
difference between the radial velocity measured here and the radial
velocity published in \cite{Nidev}.  This difference in radial
velocities offers a measure of agreement in the two radial velocity
measurements despite two different methods used to compute them and
two different sets of spectra used to measure them in the two papers.
The largest of these two uncertainty estimates, but not less than
0.030 \kmse, is listed in Tables 1 and 2 as the final estimate of the
1-sigma uncertainty for the radial velocity of each standard star.  We
adopted this floor 0.030 \kms for the stated uncertainty because this
was the uncertainty of the velocities given in \cite{Nidev}.  Any
fortuitous agreement between the current velocities and those in
Nidever et al. that happens to be smaller than 0.030 \kms could well
be spurious.  This adopted floor at 0.030 \kms prevents our quoted
measurement uncertainty from dropping lower than the level below which
we have no useful comparison with the Nidever et al. velocities.

\begin{figure} 
\includegraphics[scale = 0.6]{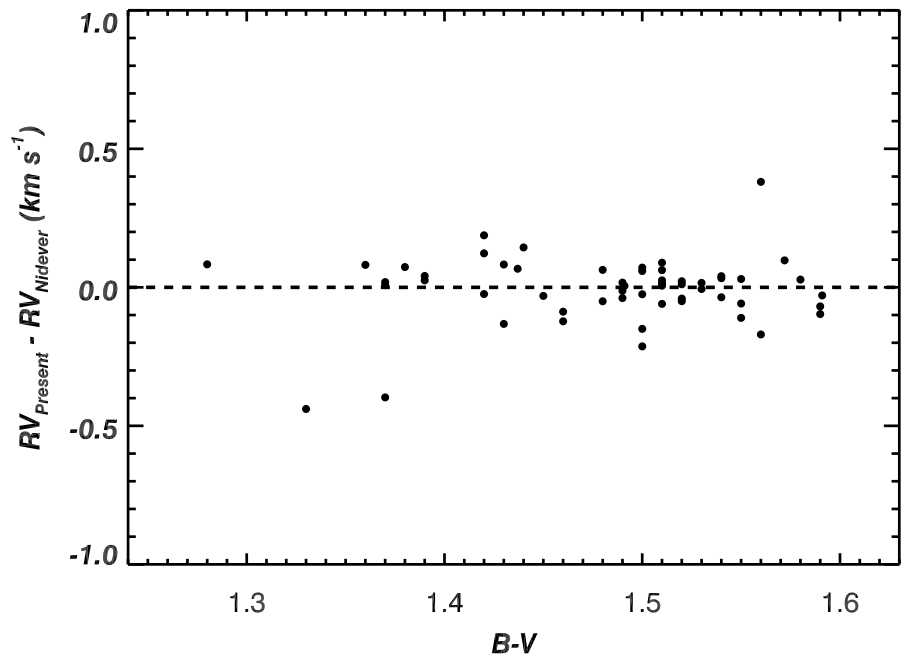} 
\caption{Difference between present velocities and those of \cite{Nidev} for all 52 M dwarfs in common.  The RMS of 0.13 \kms implies errors of no more than that magnitude for the M dwarfs in both sets.}
\label{fig:NidM}
\end{figure}

To broaden the scope of this uncertainy assessment for the standard
stars, we compared the measured radial velocities here to those in
common with \cite{Nidev} among the full set of \alln stars, not just
the standard stars.  We display the difference between our present
radial velocities and those of \cite{Nidev} in Figure \ref{fig:NidNOM}
for FGK stars and Figure \ref{fig:NidM} for the M dwarfs.  For the 428
FGK stars in common, the differences have an RMS of 0.13 \kmse.
Thus the combined errors in the present work and in Nidever et
al. amount to 0.13 \kms for the FGK stars, as described in more detail in Section 2.3.
For the 52 M dwarfs in common, the differences exhibit an RMS of 0.13
\kms (with three outliers near 0.4 \kmse) indicating the level of combined errors among M dwarfs in the
two studies.  The errors in the final velocities for the standard
stars will be smaller than quoted above for the entire set of \alln
stars because the standards typically have more observations and have
constant radial velocities by their selection.

\subsection{Radial Velocities of \alln Stars}

Table 3 reports the radial velocities of all \alln stars (including
the standards) relative to the solar system barycenter.  The same
technique that was used to determine the radial velocities of the
standard stars was used to determine the radial velocities for all
\alln stars.  In Table 3, the primary star name is given in column 1,
and the template type in column 2.  The symbol "V" represents the
Vesta spectrum (solar), and "M" represents the constructed M-dwarf
template described above.  The 3rd column gives the unweighted mean of
the Julian Dates of our observations, and the 4th column gives the
number of days between the first and last observation.  For each star,
we compute the unweighted average of all radial velocity measurements
from all spectra we obtained for that star.  The 5th column gives that
average radial velocity for the star, measured in the frame of the
barycenter of the solar system.  The 6th column gives the number of
radial velocity observations, and the 7th column gives the standard
deviation of all radial velocity measurements of that star, a measure
of both the uncertainty and of the intrinsic variation of the radial
velocities.

Examination of Table 3 shows that among the \alln stars with measured
radial velocities, some stars have only one or two velocity
measurements while others have over 30 measurements.  The time span
between the first and last spectrum is typically over a year, and
often many years.  The standard deviation of the velocities given in
the last column is a measure of the combined errors and acceleration
of the star during the time span of observations.  The median standard
deviation is 0.12 \kms, representing the uncertainty of our
individual velocity measurements, but
increased slightly by the actual velocity variations of the stars.

One measure of the 1-sigma errors of the radial velocities in Table 3
is given by the uncertainty in the mean, namely, $\sigma_{\rm RV}/N_{\rm Obs}$.
However, because of systematic errors caused by convection of $\sim$0.1 \kmse
described in Section 4, we prefer to avoid stating the formal
uncertainties that could be misinterpreted as useful uncertainties.
Also, some stars exhibit intrinsic velocity variation caused by unseen
orbiting companions, thus artifically augmenting the formal uncertainty in
the mean.

Nonetheless, examination of $\sigma_{\rm RV}$ in the last column of
Table 3 shows scatter of typically 0.15 \kms for individual
velocity measurements, serving as an upper limit to the typical errors.
Thus any values of $\sigma_{\rm RV}$ in Table 3 greater than
0.45 \kms (3 sigma) are likely ``real'', i.e. indicating actual
changes in the radial velocity of that star by an amount given by that
standard deviation on a time scale constrained by the time span of the
observations.

\begin{figure} 
\includegraphics[scale = 0.6]{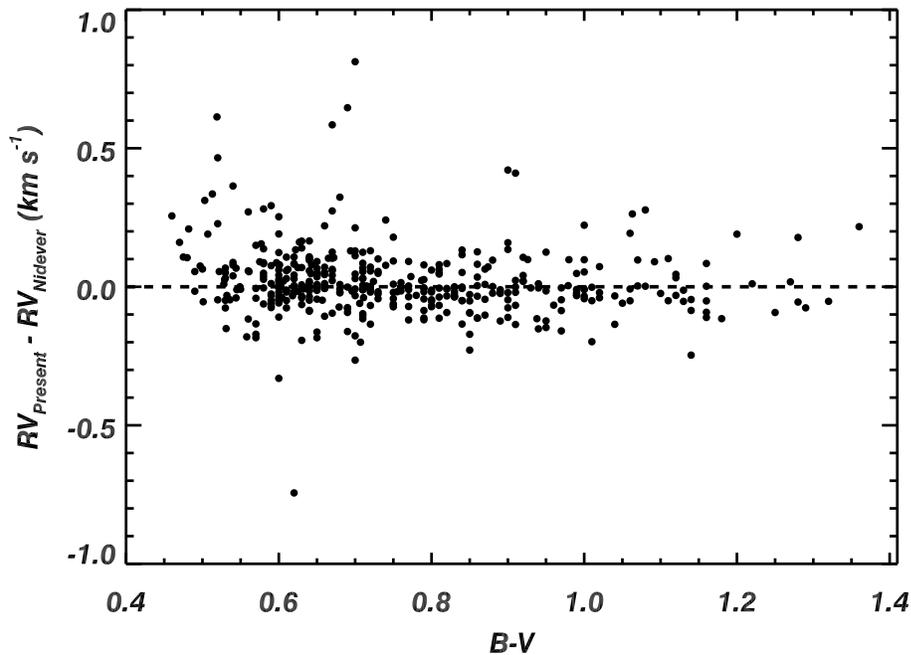} 
\caption{Difference between present velocities and those of
  \cite{Nidev} for all 428 F,G, and K stars in common plotted versus
  B-V. The scatter of 0.1 \kms demonstrates that the present
  velocities agree with those of Nidever et al. (2002) with a joint
  RMS of 0.1 \kmse, and that they share a common zero-point and
  velocity scale.  There are six stars with a velocity difference
  (present minus Nidever et al.)  larger than 0.5 \kms namely HD 87359 (+0.81 \kmse), HD 114174 (+0.58
\kmse), HD 180684 (+0.61 \kmse), HD 196201 (+0.65 \kmse),
and HD 91204 (-0.74 \kmse), and HD217165 (-2.2 \kmse).  All are
clearly binary stars as confirmed by coherent variations of over 0.5
\kms in our iodine-based {\em relative} RV measurements for each of them.}
\label{fig:NidNOM}
\end{figure}

We have compared the present velocities to those of Nidever et
al. (2002).  Nidever et al. used the iodine lines and the visible
portion of the spectrum to measure Doppler shifts, a method quite
independent of that used here.  Thus a comparison of the two sets of
the velocities for stars in common offers a method of identifying
random and systematic errors that stem from the Doppler methods
themselves.  Figure \ref{fig:NidNOM} shows the difference between the
present velocities and those of Nidever et al. (2002) as a function of
stellar color, B-V, for all 428 stars in common classified as F, G, or
K spectral type.  The plot shows that the differences in the
velocities are typically less than 0.2 \kmse, with an RMS of the differences of 0.13 \kmse,
and there is no evidence of a dependence on stellar color.  (We
removed HD 217165 from the RMS calculation, which is a binary star.)
This suggests that the accuracy and zero-points of the present and Nidever
et al. velocities are similar within $\sim$ 0.13 \kms (RMS).

However, Figure \ref{fig:NidNOM} reveals six stars (with one off scale) for which the
difference between present and Nidever et al. velocities are over 0.5
\kmse ($>$3 sigma differences).  These stars are HD 87359 (+0.81 \kmse), HD 114174 (+0.58 \kmse),
HD 180684 (+0.61 \kmse), HD 196201 (+0.65 \kmse), HD 91204 (-0.74 \kmse), and HD217165 (-2.2 \kmse). 
Examination of the iodine-based relative velocities (precise
to 0.002 \kms RMS) for these six stars reveals all of them to exhibit
long-term trends of velocity of over 0.5 \kmse.  These are
certainly long period binary stars.  The difference between the
present velocities and those of Nidever et al. (2002) is simply due to
the orbital motion that has occurred since the spectra were taken for
the work of Nidever et al. (prior to 2002) and those here that were
taken after 2004 June.  Thus, the present velocities offer a sieve for
binary stars,

Figure \ref{fig:NidM} shows a similar comparison of present velocities
and those of Nidever et al. (2002) for the 52 M dwarfs in common.  The
RMS of the differences of 0.13 \kms indicates larger errors for the M dwarfs
than for the FGK-type stars.  This error is reminiscent of that
seen in Figure \ref{fig:compare_stef} for which the difference between
the present velocities and those of Udry et al. (1999a) among the M
dwarfs was 0.15 \kmse.  These metrics suggests that the M dwarf
velocities in general, from all surveys, remain uncertain at the level
of 0.2 \kms (RMS) and harbor uncertain zero points at the level of
0.15 \kmse.

\begin{figure} 
\includegraphics[scale = 0.6]{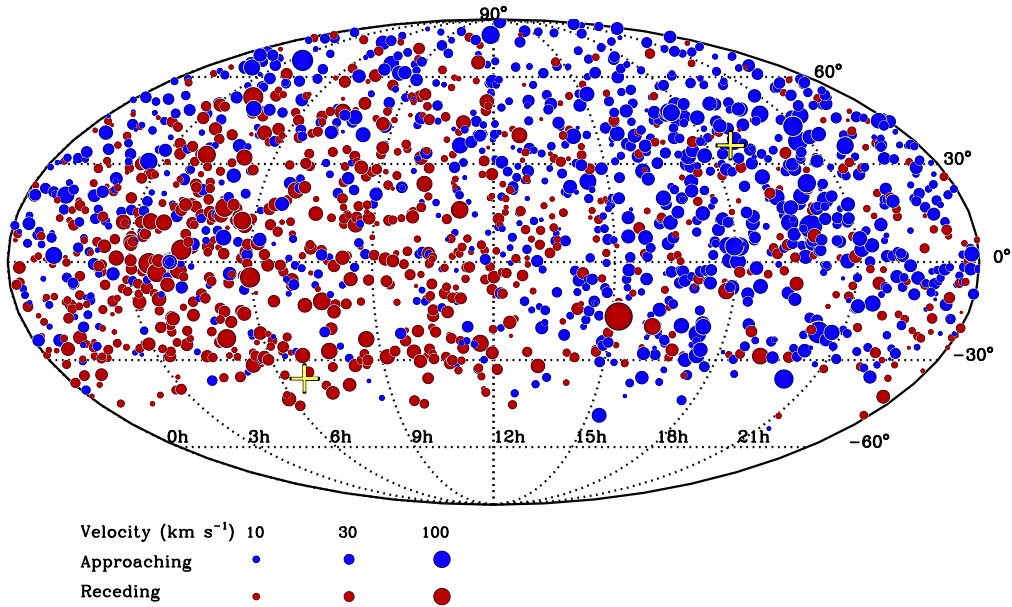} 

\caption{Location on the celestial sphere and the radial velocities relative to the barycenter of
  the Solar System of the \alln FGKM stars in the Solar neighborhood
  presented here.  The radial velocities are color-coded, with blue and red
  corresponding to approaching and receding stars, respectively,
  relative to the solar system barycenter. The diameter of the dot is
  proportional to the square root of the absolute value of the radial
  velocity (see the key).  The coordinates are equatorial in a
  Mollweide projection.   Stars are located at all RA and northward of Dec = -50
  deg.  The large number of stars near the equator, as well as in the
  north and south, offers a variety of different spectral types as
  reference stars accessible from ground-based observatories
  worldwide. The Solar Apex is marked with a cross
  at RA = 18 hr 40 min, DEC = +35 deg 58 min,
  indicating the direction of the Sun's velocity vector relative to
  nearby G dwarfs \citep{Abad03}. The solar antapex
  is also marked.  The 
  large red dot at RA$\sim$15hr 10 min and DEC$\sim$-16 deg represents two
  stars, HD 134439, HD 134440 (both metal poor halo stars)
 receding at  309.4 and 310.0  \kmse, respectively, approaching the escape
  velocity (550 \kmse) of the Milky Way Galaxy. The star with the
  greatest blueshift is HD 188510 (RA = 19 55 9.6, DEC = +10 44 27), approaching at 192.6 \kmse.}
\label{fig:location} 
\end{figure}

Figure \ref{fig:location} shows the location of the stars in
equatorial coordinates in a Mollweide projection on
the sky.   The broad distribution at all RA, and northward of DEC =
-50 offers a set of secondary standard stars.    The dots are
color-coded with blue representing stars approaching and red
representing stars receding from the barycenter of the solar system.
The size of the dots is proportional to the square root of the
absolute value of the radial velocity, an arbitrary functional form
for ease in display.  The solar apex is shown as a cross, the
direction of the motion of the sun relative to the G dwarfs in the
solar neighborhood \citep{Abad03}.  Analysis of such
all-sky measurements of Doppler shifts can, in principle and after
removal of the solar apex motion, reveal effects from gravitational
redshift including tests of general relativity \citep{Hentschel94}.

 Those stars exhibiting a standard deviation of their measured velocities less than 0.1 \kms 
as listed in Table 3 (last column) and having a time span of
observations over a few years constitute secondary standard
stars.  Their lack of radial velocity variation above 0.1
\kms during several years indicates a constant velocity suitable for
many purposes.   In contrast, the standard stars listed in Tables 1
and 2 met a higher standard of constant radial velocity within 0.1
\kms during a time span of a full 10 years and all of them also
exhibted precise radial velocities (using iodine as wavelength
reference) constant to within 0.025 \kmse, thereby ensuring their
integrity as standard stars.
 
\section{Binary Stars}

We compared our radial velocities with those previously published,
noting a subset of stars that show differences of over 2 \kmse,
indicating likely binary stars.  We made great use of the Pulkovo
Catalog of Radial Velocities \citep{Gontcharov06}.  The Pulkovo
catalog gives the weighted mean absolute velocities for over 35000
Hipparcos stars drawn from over 200 publications.  Despite the
inhomogeneous sources, the median accuracy of the final radial
velocities in the Pulkovo Catalog is 0.7 \kms, adequate to identify
binaries in comparison with the absolute velocities presented here.
The times of observations from the Pulkovo Catalog were typically
10-30 years ago, offering a time difference of typically over 10 years
between those measurements and the radial velocities presented here.
Thus, binary stars with periods over a decade can be
identified. Thanks are due to Charles Francis and Erik Anderson for
their critical evaluation of the velocities in this work compared to
those in the Pulkovo Catalog of Radial Velocities.

Among the \alln stars reported here in Table 3, we identified those
having a difference between the present velocities and those in the
Pulkovo Catalog of more the 3 $\sigma$, i.e. 2 \kms.  Velocity
differences of over 2 \kms offer a sign, but not convincing evidence,
of long term binary motion with orbital periods over a year.  For
binaries with periods of between a year and several decades, the
velocity variation will be many \kms on time scales of a decade,
allowing some of them to be detected.

Table 4 gives a list of the stars showing differences of over 2 \kms
between the present and Pulkovo velocities, indicating a possible
binary. The first and second columns give the HD and Hipparcos
identities of the star.  The third column gives the radial velocity
from the present work, and the fourth column give the radial velocity
from the Pulkovo Catalog \citep{Gontcharov06}.  For each of the stars
in Table 4, we examined the relative, precise iodine-based radial
velocities (precision of $\sim$2 \mse) to detect any obvious velocity
variations.  Indeed, for many of the stars in Table 4, the precise,
iodine-based RVs reveal large velocity variations of over 1 \kms,
confirming the binary nature. For them, we note the measured time
derivative of the variation of precise radial velocities ("PRV var")
in the last column of Table 4 under "Comments".  For the remaining
stars in Table 4, we do not have precise relative RVs, and must rely
on the difference between the present and Pulkovo velocities as the
indicator of a binary star.  Certainly a few of these entries may be
false binaries, due to unavoidable errors in the Pulkovo compilation.
But we suspect that the vast majority of the stars in Table 4 are
actual binaries, and we are alerting the community to this
likelihood. In addition, we discovered several double-line
spectroscopic binaries among our target stars, so indicated in the
Comments in Table 4.

\section{Velocity Zero Point}

The present velocities share, by construction, the zero-point of the velocity scale with
that of \citet{Nidev}.  The Nidever zero-point in velocity was
determined by using spectra of both the day sky and of the asteroid,
Vesta, yielding a zero-point accurate to within 0.01 \kms for G2V stars.
The present velocities have a similarly accurate zero-point for G2V
stars.

However one must consider the effects of general relativistic gravitational
redshifts upon departure of the light from the star (but we do
include the general relativistic blueshift caused by entry of light into the potential wells of the solar
system, an effect of only 0.003 \kmse).  One must also consider the
``convective blueshift'' caused by the hydrodynamic effects in the
photospheres of FGKM stars \citep{Dravins08}.  We emphasize that our
present velocities were constructed to have the correct velocity for
the Sun and Vesta, thus automatically accounting for gravitational
redshift and convective blueshift for G2V stars.
Here we estimate these
two effects on the velocity zero-point as a function of stellar mass
along the main sequence.

The gravitational redshift of light upon departure from stars is $K$ =
0.635($M/R$) where K is given in \kms and M and R are the stellar mass
and radius given in solar units.  As $R$ is nearly proportional to $M$
along the main sequence, the gravitational redshift varies little
among the main sequence stars, and remains $\sim$0.6 \kms for FGKM stars.

But the convective velocities decrease substantially
for the lower mass stars that have lower luminosities, requiring lower
convective velocities to carry the energy.  Scaling the
convective energy transport with stellar mass suggests that
convective velocities will vary linearly with stellar mass.  Indeed,
the RV ``jitter'' decreases from 2 \ms for G dwarfs to less than 1 \ms
for M dwarfs, in part caused by the decrease in sub-photospheric
convective hydrodynamics, not necessarily due to spots.
We note that convective blueshift depends on the technique used to
measure radial velocity because it stems from a net displacement and
distortion of the absorption line profiles.  These displacements and
shapes of the lines arise from the integrated velocity field with
depth in the photosphere, implying that each radial velocity technique
with its particular set of absorption lines
will sample a different portion of that velocity field.  Given this
physical situation, it is noteworthy that the present velocities and
those of Nidever et al. (2002) show negligible discrepancies as they
sample the near-IR and green/optical portions of the spectrum,
respectively.

One may anticipate that M dwarfs of
$\sim$0.5 solar masses will suffer a convective blueshift that is only
half that of solar type stars, and hence half that necessary to cancel
the gravitational redshift.  This suggests that the radial velocities
of M dwarfs presented here may suffer from a net surplus of
gravitational redshift compared to convective blueshift of $\sim$0.3
\kmse. 

To quantify this imbalance of convective blueshift against
gravitational redshift, \citet{Nidev} draw from hydrodynamic models of
stellar atmospheres of \citet{Dravins} to estimate the resulting
systematic errors. Based on them and on computed gravitational
redshifts, we estimate that our present radial velocities are too low
by $\sim$0.56 \kms for F5V stars.  For those stars convective
blueshift causes a greater blueshift than the gravitational redshift.
For G2V stars, our present radial velocities have a zero-point
accurate to within 0.01 \kmse, by construction (using the Sun and
Vesta).  For K0V and M0V stars, our present velocities are probably
too high by $\sim$0.15 and 0.30 \kmse, respectively.  

We caution that the asymmetries in absorption lines leading to
convective blueshift vary from line to line depending on the velocity
fields at their depth of formation, with variations of $\sim$0.1
km/s.  The asymmetries also vary with time during a magnetic cycle as
the surface fields influence the convective flow patterns.  Moreover,
the convective blueshift will be a function of spectral resolution and
of the algorithm used to measure it, i.e. cross-correlation or other,
that implicity apply weights along the line profile.

{\em Thus, to
  obtain kinematically robust measures of radial velocity, d$r$/d$t$,
  we recommend applying the corrections listed above, to the velocities in Tables 1, 2, and 3, i.e. adding
  0.56 \kms to our velocities of F5V stars, zero for G2V, subtracting
  0.15 \kms for K0V, and subtracting 0.3 \kms for M0V.}  A useful
linear relation that approximately represents the correction (in \kmse) to be applied is: 

$$V_{\rm Corr} = 1.3\times10^{-4} (T_{\rm eff} - 5780K)$$

This correction to our radial velocities is pinned to the zero-point
established by the spectra of the Sun and asteroids, and applies
only to main sequence stars.

We check the velocity zero-point assessment described above, as
follows.  A careful assessment of gravitational redshifts is provided
by \citet{Pasquini2011}.  They compared the radial velocities of main
sequence stars and giants within the open cluster, M67, expecting to
find a larger gravitational redshift from the main sequence stars due
to their smaller radii.  Remarkably, their radial velocities of main
sequence stars and giants showed no difference in systemtic velocities
of the two stellar populations.  \citet{Pasquini2011} find an upper
limit of 0.1 \kms in the net difference in the systemic radial
velocities between main sequence stars and giants.  This lack of RV
difference indicates that the spectral lines in main sequence stars
are sufficiently blueshifted, compared to those in giant stars (that
suffer only a small gravitational redshift due to their large radii),
such that the gravitational redshift and convective blueshift nearly
cancel each other in FGK main sequence stars, at the level of 0.1
\kmse.  This cancellation provides some assurance that our velocity
zero-point, which is forced to be zero for G2V stars, is not highly
sensitive to changes in convective blueshift along the main sequence.

We further check our velocity zero-point by comparison with Center for
Astrophysics (CfA) radial velocity results that targeted asteroids
having known ephemerides to establish the instantaneous velocity
vectors relative to the observatory \citep{Latham02}.  This effort is
similar to that employed by \citet{Nidev} who used Vesta to set their
zero-point.  The CfA group finds that their native radial velocities
require a correction of +0.139 \kms to achieve agreement with the
actual dynamical orbital velocities of the asteroids \citep{Latham02}.

This asteroid-derived correction of +0.139 \kms to the CfA radial
velocities may be combined with the measured zero-point difference
between the present radial velocities and those from the CfA.  In
Section 2.1 we noted that the velocities of the present standard stars
differed from those of the CfA \citep{Stef99}, with a zero--point
difference: $(V_{\rm present} - V_{\rm Stefanik})$ = +0.15 \kmse.  But
the CfA radial velocities should be corrected by +0.139 based on their
asteroid reference.  Doing so reduces the difference between the
present and (corrected) CfA velocity zero point to 0.15-0.139 \kms =
0.011 \kmse.  {\em Thus, the radial velocities presented here differ
  from the dynamically derived velocity zero-point at the CfA by only
  0.011 \kmse.}

The lines of evidence presented in this section indicate that the
radial velocities presented here are accurate measures of the time
rate of change of the distance of the star from the solar system
barycenter for solar type stars. In summary, our velocity zero-point
was pinned to \citet{Nidev} that stemmed from spectra of the Sun and
Vesta.  \cite{Pasquini2011} show that gravitational redshift and
convective blueshift nearly cancel for main sequence FGK stars.  The
asteroid measurements at the CfA \citep{Latham02} yield a zero-point
of the velocity scale that agrees with that here, within 0.01
\kmse. Thus, the present velocities represent the actual time rate of
change of distance, d$r$/d$t$, of the solar-type stars.  For other
spectral types, corrections to velocities should be applied for
convective blueshift, as noted above.  We caution that even after
applying such corrections, the present radial velocities may carry
systematic errors of 0.1 \kms or more, especially for spectral types
far from G2V.
  
\section{Discussion}

We have provided barycentric radial velocities with an internal
precision of $\sim$0.1 \kms for \alln stars, of which \stann are
standards.  The error estimates come from both the internal errors
found from our measurements and from the comparison with the standard
velocities of \cite{Nidev}, \cite{Stef99} and \citet{Udry99a}.  Our
absolute radial velocities were constructed to share the velocity
zero--point defined by \cite{Nidev} and apparently the resulting
velocity scale differs by only 0.063 \kms from that of
\citet{Udry99a}, by 0.15 \kms from that of \citet{Stef99}, and 0.007
\kms from that of \citet{Marcy}, which adds confidence to the zero
points of all four sets of velocities.

 The 1-sigma errors of the radial velocities in Table 3 are $\sim$0.12
 \kmse.  Any stars exhibiting a scatter among their individual
 velocity measurements, listed as $\sigma_{\rm RV}$ in the last column
 of Table 3, that is greater than 0.36 \kms represents a 3-sigma
 departure from a constant velocity.  Such scatter likely indicates
 physical changes in the radial velocity of that star by an amount
 given by that standard deviation and occurring on a time scale
 constrained by (shorter than) the time span of the observations.  In
 such cases of velocity variability, the individual radial velocities
 and their times of observation offer information about the coherence,
 if any, and the time scale of the acceleration of the star.
 Certainly long term trends and periodicities in the radial velocities
 offer information on the cause of the velocity variation and on the
 orbital or physical behavior.

Such accelerations are likely caused by gravitational forces within
a multiple star system.   Other possible causes are gravitational
forces exerted by orbiting giant planets, passing stars including
compact objects, structural pulsations in the star itself, rapid
rotation coupled with surface inhomogeneities such as starspots,
Rossiter McLaughlin effect from orbiting objects, or stochastic
surface velocities from magnetic events such as flares.

The precise barycentric radial velocities presented here may serve as
useful reference measurements for calibrations of other spectroscopic
programs.  They may be used to construct velocity metrics for studies
of the kinematics of the Galaxy or of other galaxies.  We intended for
these velocities to be useful to surveys of Galactic kinematics and
dynamics such as {\em Gaia, SDSS, RAVE, APOGEE, SkyMapper, HERMES, and
  LSST}.  They may assist in Doppler searches for long-period binary
stars.  Indeed, these velocities will help identify long period
orbiting or passing companions to the \alln stars themselves, most of
which reside within 100 pc thus making them interesting targets for
future high contrast imaging.

\acknowledgements  
  We are indebted to the University of California and NASA for
 allocation of telescope time on the Keck telescope.  We thank Charles
 Francis and Erik Anderson for a critical review of the velocities
 compared to past measurements.  We thank Guillermo (Willie) Torres
 and Dimitri Pourbaix for valuable suggestions that improved the
 manuscript.  This work benefited from valuable discussions about
 radial velocity standard stars with Dave Latham, Stephane Udry, and
 Dainis Dravins.  We
 are grateful to UCLA for hospitality during the writing of some of
 the paper. This work made use of the Exoplanet Orbit Database and the
 Exoplanet Data Explorer at www.exoplanets.org.  All 29000 spectra are
 archived and publicly available, thanks to the Keck Observatory
 Archive made possible by a NASA-funded collaboration between the NASA
 Exoplanet Science Institute and the W. M. Keck Observatory.  We
 acknowledge support by NASA grants NAG5-8299, NNX11AK04A, NSF grants
 AST95-20443 (to GWM) and AST-1109727 (to JTW), and by Sun
 Microsystems.  This research was made possible by the generous
 support from the Watson and Marilyn Albert SETI Chair fund (to GWM)
 and by generous donations from Howard and Astrid Preston. The Center
 for Exoplanets and Habitable Worlds is supported by the Pennsylvania
 State University, the Eberly College of Science, and the Pennsylvania
 Space Grant Consortium.  This research has made use of NASA's
 Astrophysics Data System and the SIMBAD database, operated at CDS,
 Strasbourg, France.  We thank R.Paul Butler and Steven Vogt for help
 making observations.  We thank the staff of the W.M Keck Observatory
 and Lick Observatory for their valuable work maintaining and
 improving the telescopes and instruments, without which the
 observations would not be possible.  We appreciate the State of
 California for its support of operations at both observatories.  We
 thank the University of California, Caltech, the W.M. Keck
 Foundation, and NASA for support that made the Keck Observatory
 possible.  We appreciate the indigenous Hawaiian people for the use
 of their sacred mountain, Mauna Kea.

\clearpage

\LongTables
\begin{deluxetable}{lllcccrrrrrrl}
\tablecaption{Radial Velocities of Standard Stars}
\tablewidth{0pt}
\tablehead{
  \multicolumn{1}{l}{HD\#} & 
  \multicolumn{1}{l}{HIP} &  
  \multicolumn{1}{l}{Gliese}&  
  \multicolumn{1}{c}{Spec.}& 
  \multicolumn{1}{c}{JD$_{\rm Init}$}&  
  \multicolumn{1}{c}{JD$_{\rm Fin}$} &  
  \multicolumn{1}{c}{$\Delta$t} & 
  \multicolumn{1}{c}{RV$_{\rm Nidever}$\tablenotemark{a}} &  
  \multicolumn{1}{c}{RV$_{\rm Present}$\tablenotemark{b}} & 
  \multicolumn{1}{c}{$\sigma_{\mathrm{RV}}$\tablenotemark{c}} & 
  \multicolumn{1}{c}{$N_{\rm Obs}$\tablenotemark{d}}  & 
  \multicolumn{1}{l}{RV$_{\rm Final}$\tablenotemark{e}}  & 
  \multicolumn{1}{l}{Unc\tablenotemark{f}}\\ 
   \multicolumn{1}{l}{}  & 
   \multicolumn{1}{l}{} & 
   \multicolumn{1}{l}{}  & 
   \multicolumn{1}{c}{Type} & 
   \multicolumn{1}{c}{(-2440000)}  & 
   \multicolumn{1}{c}{(-2440000)} & 
   \multicolumn{1}{c}{(yr)}  & 
   \multicolumn{1}{c}{(km/s)} & 
   \multicolumn{1}{c}{(km/s)} & 
   \multicolumn{1}{l}{(km/s)} & 
   \multicolumn{1}{l}{} & 
   \multicolumn{1}{l}{(km/s)} & 
   \multicolumn{1}{l}{(km/s)}     
  } 
\startdata
166 &   544 &   5 & K0 &         6960 &        15252 &           23 &     -6.537 &     -6.350 &      0.118 &           12 &     -6.444 &      0.187 \\ 
283 &   616 &   9003 & K0 &        10367 &        14457 &           11 &    -43.102 &    -43.247 &      0.077 &            7 &    -43.174 &      0.145 \\ 
3651 &   3093 &   27 & K0 &         7048 &        15251 &           22 &    -32.961 &    -32.919 &      0.108 &           63 &    -32.940 &      0.042 \\ 
3765 &   3206 &   28 & K2 &        10462 &        15252 &           13 &    -63.202 &    -63.179 &      0.119 &           62 &    -63.191 &      0.030 \\ 
4256 &   3535 &   31.4 & K2 &        10367 &        15172 &           13 &      9.460 &      9.393 &      0.104 &           49 &      9.426 &      0.067 \\ 
4628 &   3765 &   33 & K2 &         7047 &        15252 &           22 &    -10.230 &    -10.228 &      0.121 &           79 &    -10.229 &      0.030 \\ 
8389 &   6456 &   57.1 & K0 &        10367 &        15016 &           13 &     34.647 &     34.606 &      0.110 &           58 &     34.626 &      0.041 \\ 
9562 &   7276 &   59.2 & G2 &        10367 &        15230 &           13 &    -14.989 &    -14.990 &      0.132 &            7 &    -14.990 &      0.050 \\
\enddata
\tablenotetext{a}{Radial velocities from Nidever et al. (2002)}
\tablenotetext{b}{Radial velocities from the measurements made here.}
\tablenotetext{c}{Standard deviation of all RVs measured here for this star.}
\tablenotetext{d}{Number of observations (spectra) used for this star.}
\tablenotetext{e}{Unweighted average of the Nidever RV and the RV measured here.  This is the recommended RV for use as a standard star.}
\tablenotetext{f}{Uncertainty in the Final RV (see text).}
\end{deluxetable}
\clearpage

\LongTables
\begin{deluxetable}{lccccrr}
\tablecaption{Radial Velocity Standard Stars with Coordinates}
\tabletypesize{\scriptsize}
\tablewidth{0pt}
\tablehead{
  \multicolumn{1}{c}{Star Name}   &
  \multicolumn{1}{c}{RA}   &
  \multicolumn{1}{c}{DEC}   &
  \multicolumn{1}{c}{V$_{\mathrm{mag}}$}   &
  \multicolumn{1}{c}{Spectral}   &
  \multicolumn{1}{c}{Final RV}   &
  \multicolumn{1}{c}{Unc} \\
   \multicolumn{1}{l}{}   &
   \multicolumn{1}{l}{(2000)}   &
   \multicolumn{1}{l}{(2000)}   &
   \multicolumn{1}{c}{}   &
   \multicolumn{1}{c}{Type}   &
   \multicolumn{1}{c}{(km s$^{-1}$)}   &
 \multicolumn{1}{c}{(km s$^{-1}$)}
 }
 \startdata
HD 166 &            0           6      36.8 &           29           1      17.4 &       6.13 & K0 &     -6.444 &      0.187 \\ 
HD 283 &            0           7      32.5 &          -23          49       7.4 &       8.70 & K0 &    -43.174 &      0.145 \\ 
HD 3651 &            0          39      21.8 &           21          15       1.7 &       5.80 & K0 &    -32.940 &      0.042 \\ 
HD 3765 &            0          40      49.3 &           40          11      13.8 &       7.36 & K2 &    -63.191 &      0.030 \\ 
HD 4256 &            0          45       4.9 &            1          47       7.9 &       8.03 & K2 &      9.426 &      0.067 \\ 
HD 4628 &            0          48      23.0 &            5          16      50.2 &       5.75 & K2 &    -10.229 &      0.030 \\ 
HD 8389 &            1          23       2.6 &          -12          57      57.8 &       7.84 & K0 &     34.626 &      0.041 \\ 
HD 9562 &            1          33      42.8 &           -7           1      31.2 &       5.76 & G2 &    -14.990 &      0.050 \\
\enddata
\end{deluxetable}
\clearpage

\begin{deluxetable}{lrrccrr}
\tablecaption{Absolute Radial Velocities for 2046 Stars}
\tabletypesize{\scriptsize}
\tablewidth{0pt}
\tablehead{
  \multicolumn{1}{c}{Star Name}   &
  \multicolumn{1}{c}{Template\tablenotemark{a}}   &
  \multicolumn{1}{c}{$<$JD$>$}   &
  \multicolumn{1}{c}{$\Delta$T}   &
  \multicolumn{1}{c}{$<$RV$>$}   &
  \multicolumn{1}{c}{Obs}   &
  \multicolumn{1}{c}{$\sigma_{\mathrm{RV}}$} \\
   \multicolumn{1}{l}{}   &
   \multicolumn{1}{l}{}   &
   \multicolumn{1}{l}{-2450000}   &
   \multicolumn{1}{c}{(days)}   &
   \multicolumn{1}{c}{(km s$^{-1}$)}   &
   \multicolumn{1}{c}{}   &
 \multicolumn{1}{c}{(km s$^{-1}$)}
 }
 \startdata
HD 224983 &  V &         4176 &         1619 &    -17.486 &            8 &      0.136 \\
HD 225021 &  V &         4646 &          832 &     -6.225 &            7 &      0.176 \\
HD 225118 &  V &         3810 &          594 &     10.869 &            9 &      0.069 \\
HD 225261 &  V &         3585 &          692 &      7.493 &            2 &      0.022 \\
HIP 428 &  M &         4292 &         1933 &     -0.287 &           13 &      0.175 \\
HD 225213 &  M &         4290 &         1474 &     25.379 &           34 &      0.094 \\
HD 38b &  M &         4219 &         1480 &     -1.700 &           18 &      0.101 \\
HD 38a &  M &         4172 &         1482 &      1.654 &           10 &      0.142 \\
\enddata
\tablenotetext{a}{We used the Vesta spectrum as a template when V is listed, and a spectrum for HIP 80824 as a template when M is listed.}
\tablenotetext{b}{2MASS IDs.}
\end{deluxetable}
\clearpage

\LongTables
\begin{deluxetable}{llccl}
\tablecaption{Radial Velocities Different by over 2 \kms from Pulkova}
\tabletypesize{\scriptsize}
\tablewidth{0pt}
\tablehead{
  \multicolumn{1}{c}{HD}   &
  \multicolumn{1}{c}{HIP}   &
  \multicolumn{1}{c}{Present}   &
  \multicolumn{1}{c}{Pulkova}   &
  \multicolumn{1}{c}{Comment}   
 }
 \startdata
1293 & 1365  & 45.14 & 42.9 &   PRV var: +0.080 km/s/yr \\ 
3795 & 3185  & -45.481	 & -47.4 & 	PRV var: +0.12 km/s/yr \\ 
11616 & 8872  & -11.79	 & 35.5	 & Double line, SB2: $\sim$50 km/s \\ 
11970 & 9199  & -14.377	 & -7.2	 & SB1, K=5km/s, P=25yr \\ 
---  &  13447 & -1.291	 & 14	 &     No PRV	 \\ 
27990 & 20679  & 42.593	 & 37.9 & 	No PRV \\ 
28192 & 20752  & -4.138	 & -6.2	 & No PRV \\ 
30649 &  22596 &  32.241 & 	28.7 & 	SB1 binary: K=1.5 km/s, P=24yr \\
\enddata
\end{deluxetable}
\clearpage

\clearpage 
 \begin{thebibliography}{}

\bibitem[Abad et al.(2003)]{Abad03} Abad, C., Vieira, K., Bongiovanni, A.,
  Romero, L., \& Vicente, B.\ 2003, \aap, 397, 345 

\bibitem[Beavers \& Eitter(1986)]{Beavers86} Beavers, W.~I., \& Eitter, J.~J.\ 1986, \apjs, 62, 147 

\bibitem[Burkart et al.(2012)]{Burkart2012} Burkart, J., Quataert, 
E., Arras, P., \& Weinberg, N.~N.\ 2012, \mnras, 421, 983 

\bibitem[Casagrande et al.(2011)]{Casagrande2011} Casagrande, L., 
Schoenrich, R., Asplund, M., Cassisi, S., Ramirez, I., Melendez, J., 
Bensby, T., \& Feltzing, S.\ 2011, arXiv:1103.4651 

\bibitem[Crifo et 
al.(2010)]{Crifo10} Crifo, F., Jasniewicz, G., Soubiran, C., et al.\ 2010, \aap, 524, A10 


\bibitem[Dravins(1999)]{Dravins} Dravins, D.\ 1999, IAU 
Colloq.~170: Precise Stellar Radial Velocities, 185, 268 

\bibitem[Dravins(2008)]{Dravins08} Dravins, D.\ 2008, \aap, 492, 199 

\bibitem[Edwards et al.(2006)]{Edwards2006} Edwards, R.~T.,
  Hobbs, G.~B., \& Manchester, R.~N.\ 2006, \mnras, 372, 1549 

\bibitem[Famaey et al.(2005)]{Famaey05} Famaey, B., Jorissen, A., Luri, X., et al.\ 2005, \aap, 430, 165 

\bibitem[Gilmore et al.(2012)]{Gilmore12} Gilmore, G.,  Randich, S., Asplund, M., et al.\ 2012, The Messenger, 147, 25 

\bibitem[Gontcharov(2006)]{Gontcharov06} Gontcharov, G.~A.\ 2006, 
Astronomy Letters, 32, 759 

\bibitem[Griffin et~al.(2002)]{Griffin73} Griffin, R. \& Griffin, R.  1973, Mon. Not. R. astr. Soc. 162, 243-253.

\bibitem[Griffin(1973)]{Griffin73} Griffin, R.\ 1973, \mnras, 162, 243 

\bibitem[Hearnshaw \& Scarfe(1999)]{Hearnshaw99} Hearnshaw, J.~B., \& Scarfe, C.~D.\ 1999, IAU Colloq.~170: Precise Stellar Radial Velocities, 185,  
\bibitem[Hentschel(1994)]{Hentschel94} Hentschel, K.\ 1994,  Archive for History of Exact Sciences, 47, 143 

\bibitem[Holmberg et al.(2009)]{Holmberg09} Holmberg, J., Nordstr{\"o}m, B., \& Andersen, J.\ 2009, \aap, 501, 941 

\bibitem[Johnson et al.(2011)]{Johnson11} Johnson, J.~A., 
Clanton, C., Howard, A.~W., et al.\ 2011, \apjs, 197, 26 
  
\bibitem[Kibrick et al.(2006)]{Kibrick06} Kibrick, R.~I., Clarke, 
D.~A., Deich, W.~T.~S., \& Tucker, D.\ 2006, \procspie, 6274,  

 \bibitem[Latham et al.(1991)]{Latham91} Latham, D.~W., Stefanik, 
R., Torres, G., Davis, R.~J., 
\& Mazeh, T.\ 1991, Montreal International Astronautical Federation Congress,  

\bibitem[Latham et al.(2002)]{Latham02} Latham, D.~W.,
  Stefanik, R.~P., Torres, G., et al.\ 2002, \aj, 124, 1144 

\bibitem[Lindegren \& Dravins(2003)]{Lindegren03} Lindegren, L., \& Dravins,
  D.\ 2003, \aap, 401, 1185 

\bibitem[Marcy et al.(1987)]{Marcy} Marcy, G.~W., Lindsay, 
V., \& Wilson, K.\ 1987, \pasp, 99, 490 

\bibitem[Marcy \& Butler(1992)]{Marcy92} Marcy, G.~W., \& Butler, R.~P.\ 1992, \pasp, 104, 270 

\bibitem[Marcy et al.(2008)]{Marcy08} Marcy, G.~W., Butler, 
R.~P., Vogt, S.~S., et al.\ 2008, Physica Scripta Volume T, 130, 014001 

\bibitem[Mayor \& Maurice(1985)]{Mayor85} Mayor, M., \& Maurice, E.\ 1985, Stellar Radial Velocities, 299 

\bibitem[Nidever et~al.(2002)]{Nidev} Nidever, D.~L., Marcy, 
G.~W., Butler, R.~P., Fischer, D.~A., \& Vogt, S.~S.\ 2002, \apjs, 141, 503

\bibitem[Nordlund(2008)]{Nordlund09} Nordlund, {\AA}.\ 2008, 
Physica Scripta Volume T, 133, 014002 

\bibitem[Nordstr{\"o}m et 
al.(2004)]{Nordstrom04} Nordstr{\"o}m, B., et al.\ 2004, \aap, 418, 989 

\bibitem[Pasquini et 
al.(2011)]{Pasquini2011} Pasquini, L., Melo, C., Chavero, C., et al.\ 2011, \aap, 526, A127 

\bibitem[Pourbaix et al.(2002)]{Pourbaix02} Pourbaix, D., et al.\ 2002, \aap, 386, 280 

\bibitem[Ram{\'{\i}}rez et al.(2010)]{Ramirez10} Ram{\'{\i}}rez, 
I., Collet, R., Lambert, D.~L., Allende Prieto, C., 
\& Asplund, M.\ 2010, \apjl, 725, L223 

\bibitem[Scarfe et al.(1990)]{Scarfe90} Scarfe, C.~D., Batten, 
A.~H., 
\& Fletcher, J.~M.\ 1990, Publications of the Dominion Astrophysical Observatory Victoria, 18, 21 

\bibitem[Schoenrich et al.(2010)]{Schoenrich2010} Schoenrich, R., 
Asplund, M., \& Casagrande, L.\ 2010, arXiv:1012.0842 


 \bibitem[Stefanik et~al.(1999)]{Stef99} Stefanik, R.~P., Latham, D.~W., \& Torres, G. 
1999, IAU Coll 170, 354 

\bibitem[Udry et~al.(1999a)]{Udry99a} Udry, S., Mayor, M., \& Queloz, D. 1999a, 
IAU Coll 170, 367 

\bibitem[Udry et~al.(1999b)]{Udry99b} Udry, S., et~al. 1999b, IAU Coll 170, 383

\bibitem[Welsh et al.(2011)]{Welsh2011} Welsh, W.~F., Orosz, 
J.~A., Aerts, C., et al.\ 2011, \apjs, 197, 4 

\bibitem[Wilson(1953)]{Wilson53} Wilson, R.~E.\ 1953, Carnegie 
Institute Washington D.C.~Publication, 0 

\bibitem[Wilson et al.(2011)]{Wilson2011} Wilson, M.~L., et al.\ 
2011, \mnras, 260 

\end{thebibliography}
\end{document}